\documentclass{emulateapj}

\usepackage{amsmath}
\usepackage{amssymb}
\usepackage{graphicx}
\usepackage{color}
\usepackage{natbib}
\usepackage{epsfig}

\def\gtaprx {\lower .1ex\hbox{\rlap{\raise .6ex\hbox{\hskip .3ex
	{\ifmmode{\scriptscriptstyle >}\else
		{$\scriptscriptstyle >$}\fi}}}
	\kern -.4ex{\ifmmode{\scriptscriptstyle \sim}\else
		{$\scriptscriptstyle\sim$}\fi}}}
\def\ltaprx {\lower .1ex\hbox{\rlap{\raise .6ex\hbox{\hskip .3ex
	{\ifmmode{\scriptscriptstyle <}\else
		{$\scriptscriptstyle <$}\fi}}}
	\kern -.4ex{\ifmmode{\scriptscriptstyle \sim}\else
		{$\scriptscriptstyle\sim$}\fi}}}

\newcommand{\cutt}[1]{\textcolor{blue}{}}

\newcommand{\Ms}{{\ensuremath{{M}_{\odot} }}}
\newcommand{\Zs}{\ensuremath{Z_\odot}}

\newcommand{\Ni}{{\ensuremath{^{56}\mathrm{Ni}}}}

\newcommand{\HII}{{\ion{H}{2}}}

\begin{document}

\title{The Biggest Explosions in the Universe. II.}

\author{Daniel J. Whalen\altaffilmark{1,2}, Jarrett L. Johnson\altaffilmark{3}, 
Joseph Smidt\altaffilmark{1}, Alexander Heger\altaffilmark{4}, Wesley 
Even\altaffilmark{5} and Chris L. Fryer\altaffilmark{5}}

\altaffiltext{1}{T-2, Los Alamos National Laboratory, Los Alamos, NM 87545}

\altaffiltext{2}{Universit\"{a}t Heidelberg, Zentrum f\"{u}r Astronomie, Institut f\"{u}r 
Theoretische Astrophysik, Albert-Ueberle-Str. 2, 69120 Heidelberg, Germany}

\altaffiltext{3}{XTD-6, Los Alamos National Laboratory, Los Alamos, NM 87545}

\altaffiltext{4}{Monash Centre for Astrophysics, Monash University, Victoria, 
3800, Australia}

\altaffiltext{5}{CCS-2, Los Alamos National Laboratory, Los Alamos, NM 87545}

\begin{abstract}

One of the leading contenders for the origin of supermassive black holes at 
$z \gtrsim$ 7 is catastrophic baryon collapse in atomically-cooled halos at $z
\sim$ 15.  In this scenario, a few protogalaxies form in the presence of strong 
Lyman-Werner UV backgrounds that quench H$_2$ formation in their
constituent halos, preventing them from forming stars or blowing heavy 
elements into the intergalactic medium prior to formation.  At masses of 10$^
8$ \Ms\ and virial temperatures of 10$^4$ K, gas in these halos rapidly cools 
by H lines, in some cases forming 10$^4$ - 10$^6$ \Ms\ Pop III stars and, a 
short time later, the seeds of supermassive black holes.  Instead of collapsing 
directly to black holes some of these stars died in the most energetic 
thermonuclear explosions in the universe. We have modeled the explosions 
of such stars in the dense cores of line-cooled protogalaxies in the presence 
of cosmological flows.  In stark contrast to the explosions in diffuse regions in 
previous simulations, these SNe briefly engulf the protogalaxy but then 
collapse back into its dark matter potential.  Fallback drives turbulence that 
efficiently distributes metals throughout the interior of the halo and fuels the 
rapid growth of nascent black holes at its center.  The accompanying starburst 
and x-ray emission from these line-cooled galaxies easily distinguish them from 
more slowly evolving neighbors and might reveal the birthplaces of supermassive 
black holes on the sky.

\vspace{0.1in}

\end{abstract}


\section{Introduction}

The existence of supermassive black holes (SMBHs) in galaxies at $z \gtrsim$ 7
\citep{fan03,wm03,fan06,mort11} poses one of the greatest challenges to the 
paradigm of hierarchical structure formation \citep{bl03,begel06,jb07b,brmvol08,
lfh09,th09,pan12b,schl13,choi13}.  One of the leading contenders for the origin of 
SMBHs is catastrophic baryon collapse in atomically cooled halos at $z \sim$ 10 - 
15 \citep{wta08,rh09,sbh10,whb11,latif13c,latif13a}.  In this picture, a few primeval 
galaxies \citep{jgb08,get08,jlj09,get10,jeon11,pmb11,wise12,pmb12} formed in 
strong Lyman-Werner (LW) UV backgrounds that photodissociated all their H$_2$, 
preventing them from forming primordial stars prior to assembly \citep{bcl99,abn00,
abn02, bcl02,nu01,on07,on08,wa07,y08,turk09,stacy10,sm11,clark11,get11,hos11,
get12,susa13}.  Unlike other primitive galaxies, which began to reionize \citep{
wan04,ket04,abs06,awb07,wa08a} and chemically enrich the IGM \citep{mbh03,
ss07,bsmith09,jet09b,jw11,ritt12,chiaki12,ss13}, line-cooled protogalaxies 
remained inert until reaching masses of $\sim$ 10$^8$ \Ms\ and virial temperatures 
of $\sim$ 10$^4$ K.

These temperatures activated rapid cooling in the halo by H lines, triggering 
massive baryon collapse with central infall rates of up to 0.1 - 1 \Ms\ yr$^{-1}$, 
1000 times those that formed the first stars in less massive halos at higher 
redshifts.  Such accretion rates were capable of building up very massive gas 
clumps in short times, most of which collapsed directly to 10$^4$ - 10$^5$ \Ms\ 
BHs.  The creation of SMBH seeds by direct baryon collapse has gained ground 
in recent years because it has been shown that Population III (Pop III) BH seeds 
cannot achieve the rapid growth needed to reach 10$^9$ \Ms\ by $z \sim$ 7 
\citep{milos09,awa09,pm11,mort11,pm12,wf12,pm13,jet13}.  It was originally 
thought that LW fluxes capable of fully suppressing Pop III star formation were 
rare, and that this might explain why so few SMBHs have been found at $z 
\gtrsim$ 6, but it is now known that such environments may have been far more 
common in the early universe than previously believed \citep{dijkstra08,jlj12a,
agarw12,petri12}.

In some cases, massive clumps in line-cooled halos formed stable stars rather 
than collapsing directly to BHs \citep{iben63,fh64,fowler66,af72a,af72b,bet84,
fuller86,begel10}.  Some of these stars have now been shown to die in 
extremely energetic thermonuclear explosions \citep{montero12,heg13} that  
will be visible in both deep-field and all-sky NIR surveys at $z \gtrsim$ 10 by 
the \textit{James Webb Space Telescope} (\textit{JWST}), \textit{Euclid}, the 
Wide-Field Infrared Survey Telescope (WFIRST), and the Wide-Field Imaging 
Surveyor for High-Redshift (WISH) \citep{wet12d} \citep[for other work on Pop 
III SN light curves, see][]{sc05,fwf10,kasen11,pan12a,hum12,det12,wet12a,
mw12,wet12b,wet12c,wet12e,ds13}.  A 55,500 \Ms\ SN in a diffuse 
environment in its host protogalaxy has been studied by \citet{jet13a}.  For 
explosions in low densities, like those of an \HII\ region created by the star, 
they find that the SN expels all the baryons from the halo to radii of $\gtrsim$ 
10 kpc.  In this scenario, metals from the SN can enrich nearby protogalaxies 
before later falling back into the halo on timescales of 50 - 70 Myr.

If the explosion instead occurs in the heavy infall that formed the star, the SN
rapidly loses energy to bremsstrahlung and line emission and expands to at
most 1 kpc, the virial radius of the halo \citep{wet13a} \citep[see also][]{ky05,
wet08a,ds11a,vas12}.  Ejecta from the SN briefly displaces the outer layers of 
the protogalaxy but then recollapses in a spectacular bout of fallback that was 
inferred to rapidly distribute heavy elements throughout its interior and fuel the 
growth of massive SMBH seeds at its center.  However, the extremely short
X-ray and line cooling timescales in the dense media plowed up by the SN 
limited these simulations to one dimension (1D), so they could not follow the 
evolution of the protogalaxy during expansion and fallback or determine how 
cosmological flows later mixed its interior with metals. 

However, it is possible to initialize a three-dimensional (3D) cosmological 
simulation with 1D blast profiles from \citet{wet13a} at intermediate times, 
after cooling timescales in the shock have become large but before it has 
reached large radii or departed from spherical symmetry.  The explosion 
can then be evolved in the protogalaxy in the presence of cosmological 
flows to determine its eventual fate, having properly determined its energy 
and momentum losses over all spatial scales.  We have now performed 
such a simulation in the GADGET code with a supermassive SN profile 
taken from the ZEUS-MP calculations of \citet{wet13a}.  We describe our 
numerical simulation in Section 2.  The evolution of the SN in the presence
of cosmological flows is examined in Section 3, where we also study the 
metallicity of the halo over time. We then discuss the potential for triggered 
starbursts and rapid central BH growth in the protogalaxy and conclude in 
Section 4.

\section{Numerical Method}

\begin{figure}
\epsscale{1.22}
\plotone{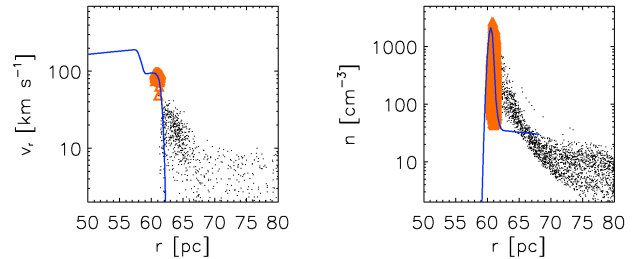} 
\caption{GADGET fit to the supermassive SN profiles from ZEUS-MP.  Left:
radial velocity.  Right: number density.  The blue curves denote the ZEUS-MP
profiles and the orange particles and triangles trace the SPH particle 
distribution assignedto these profiles.  The black points are the SPH particles 
representing the protogalaxy and cosmological flows.}
\vspace{0.1in}
\label{fig:setup}
\end{figure}

Our calculation proceeds in three stages.  First, the 55,500 \Ms\ Pop III star is 
evolved from the beginning of the main sequence to central collapse and then 
explosion in the \textit{Kepler} \citep{Weaver1978,Woosley2002} and RAGE 
codes \citep{rage,fet12}, out to 2.9 $\times$ 10$^6$ s.  The energy of the SN 
is 7.7 $\times$ 10$^{54}$ erg.  At this time the shock is at 4 $\times$ 10$^{15}
$ cm, past breakout from the surface of the star but at a radius at which it has 
not swept up much mass.  The SN is then initialized in ZEUS-MP \citep{wn06,
wn08b,wn08a} in the spherically averaged 1D profile of the halo into which it is 
later mapped in GADGET.  The density of the halo at the radius of the shock is 
$\sim$ 10$^{11}$ cm$^{-3}$ \citep[see Fig.~3 of][]{wet13a}. The SN is evolved 
out to 60 pc (at 2.54 $\times$ 10$^5$ yr) and then both it and the gas it has 
swept up are ported to GADGET.  These first two steps are described in detail 
in \citet{wet12d} and \citet{wet13a}.  

In \citet{jet13a} the SN was initialized in a diffuse $r^{-2}$ wind envelope whose
density was far lower than that of the undisturbed halo in ZEUS-MP \citep[see 
Fig.~3 of][]{wet12d}.  This wind approximated an \HII\ region created by the star.
The SN was evolved out to 6 pc in RAGE and then mapped into the center of the 
protogalaxy in GADGET. Our procedure ensures that the heavy energy losses of 
the SN in a dense protogalactic core are properly taken into account, and that the 
explosion is initialized with the proper energy and momentum in our cosmological 
simulation.     

\begin{figure*}
\plotone{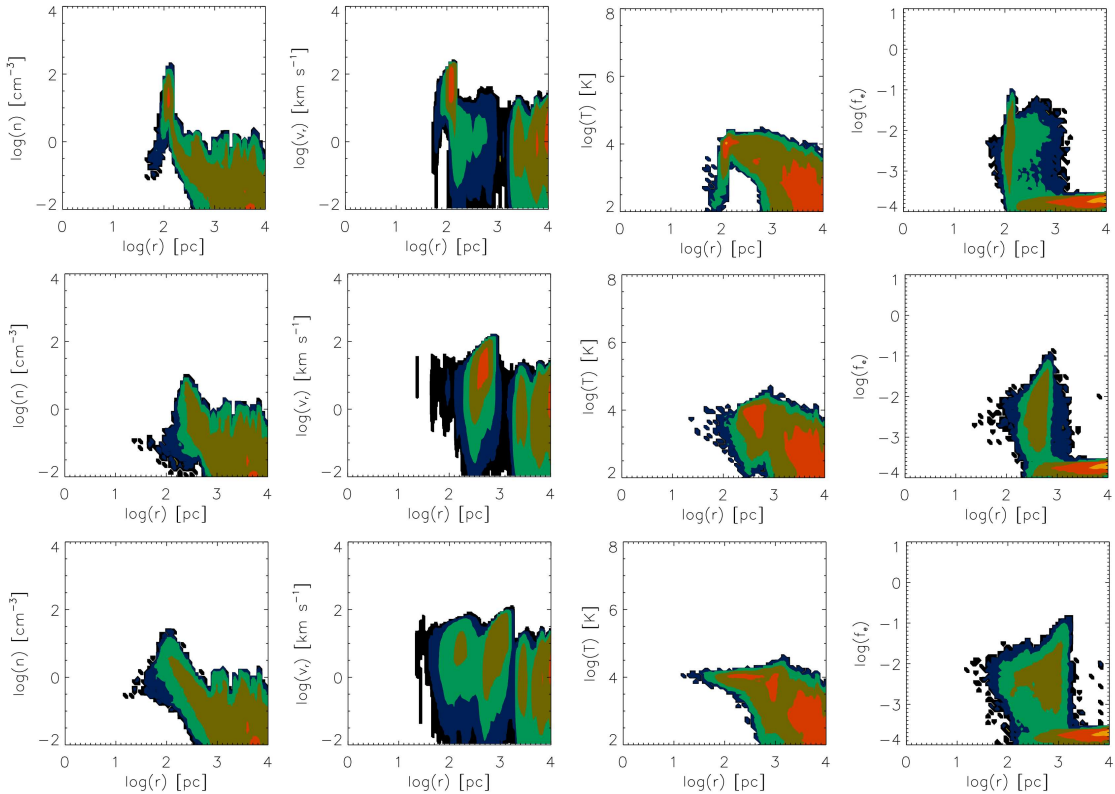} 
\caption{Protogalactic baryons plowed up by the supermassive SN at 1 Myr 
(\textit{top}), 10 Myr (\textit{center}) and 25 Myr (\textit{bottom}).  Left to right:  
H number density, radial gas velocity, temperature and free electron fraction, 
all as a function of radius from the center of the halo.  Mass fractions in the 
gas change by an order of magnitude across the contours in the plots.}
\vspace{0.1in}
\label{fig:gas}
\end{figure*}

\subsection{GADGET Model}

Our fiducial protogalaxy is the 4 $\times$ 10$^7$ \Ms\ atomically-cooled halo 
from \citet{jet13a}, which forms in a 1 Mpc$^3$ (comoving) simulation volume 
in GADGET \citep{gadget1,gadget2}.  This simulation was evolved from $z =$ 
100 down to $z \sim$ 15 in a uniform LW UV background that sterilized the 
halo of H$_2$ and prevented stars from forming in any of its constituent halos.  
The simulation that produced the protogalaxy is described in \citet{jlj11a}.  We 
map the (Eulerian) ZEUS-MP SN profile into the (Lagrangian) smoothed-particle 
hydrodynamics (SPH) GADGET model by assigning the central 11,373 SPH 
particles to the ejecta in a manner that reproduces the radial velocity and 
number density of the blast.  We show our fit to the ZEUS-MP profile in 
Fig.~\ref{fig:setup}.  The SN expels 23,000 \Ms\ of metals \citep[and probably 
molecules and dust;][]{cl08,cd09,cd10,dc11,gall11} into the galaxy.  At this 
stage it is essentially a free expansion.  

We initialize the shock at 60 pc in GADGET for two reasons.  First, the shock 
must expand and cool to temperatures at which its chemistry and cooling times 
are long enough to evolve it in a 3D cosmological simulation in a reasonable 
time.  It cannot be initialized with too large a radius because our GADGET 
simulation might then exclude asymmetries that can arise in the shock at earlier 
times.  We found that 60 pc \citep[in contrast to 6 pc in][]{jet13a} satisfied both 
requirements.  This radius was also chosen to minimize departures from mass 
conservation when assigning the SPH particle distribution to the high-resolution 
ZEUS-MP SN profile.Ê There was a fair amount of ejecta and swept-up gas in 
ZEUS-MP at radii below what could have been resolved by our GADGET model
at early times.  Allowing the SN to grow to $\sim$ 60 pc in ZEUS-MP enabled us 
to preserve its mass to within 20\% in GADGET.  Smaller errors could have 
been achieved with larger radii but at the cost of missing asymmetries in the 
shock due to inhomogeneities in the cosmological density field.  We obtained 
even better agreement in mass conservation at just 6 pc in \citet{jet13a} 
because the explosion was in lower densities that could be adequately resolved 
at smaller radii by the SPH particle distribution.  

\subsection{Nonequilibrium H/He Gas Chemistry}

We evolve mass fractions for H, H$^+$, He, He$^+$, He$^{++}$, H$^-$, H$^{+}
_{2}$, H$_{2}$, e$^-$, D, D$^+$ and HD with the 42 reaction rate network  
described in \citet{jb07a} to tally energy losses in the remnant as it sweeps up 
and heats baryons.  Cooling due to collisional excitation and ionization of H and 
He, recombinations, inverse Compton (IC) scattering from the cosmic 
microwave background (CMB), and free-free emission by bremsstrahlung 
x-rays are included. In the ZEUS-MP run H$_2$ cooling is turned off in the gas 
because high temperatures in the shock at early times destroy these fragile 
molecules.  Both H$_2$ and HD cooling are included in the GADGET run but 
are minimal because of shock heating and the LW background.  Adjustments
to LW photodissociation rates due to self-shielding by H$_2$ and H$^-$ 
photodetachment rates are taken from \citet{sbh10}.  Cooling times for these 
processes depend on temperatures and mass fractions for these species in the 
shocked gas, which in turn are governed by the energy equation and the 
nonequilibrium reaction network.      

\section{SN Evolution in the Protogalaxy}

\subsection{Dynamics and Energetics}

We show the evolution of gas in the protogalaxy as it is blown outward by the 
SN at 1, 10 and 25 Myr in Fig.~\ref{fig:gas}.  As shown in the center panels on 
the top row, the shock has cooled to $\sim$ 50,000 K and slowed to $\sim$ 300 
km s$^{-1}$ by 1 Myr, mostly because of prior energy losses to bremsstrahlung 
and H and He line emission.  Atomic lines restrict the temperature of the shock 
to 10,000 - 25,000 K thereafter (the "Balmer thermostat"), as in \citet{wet13a}.  
Consequently, free electron fractions never exceed 10\%.  At 1 Myr there are 
two components to the electron fraction: collisional ionizations in the SN shock, 
which are visible as the vertical spike in electron fraction at 100 pc, and 
ionizations in the virial shock from $\sim$ 0.5 - 1 kpc.  The propagation of the 
shock to the edge of the halo is visible as the red contours in temperature and 
radial velocity that migrate from 100 pc to 2 kpc over 25 Myr.  Similar motion is 
visible in the spike in electron fraction, which is also at 2 kpc by 25 Myr.  The 
ejecta drives most of the gas from the center of the protogalaxy at early times, 
as shown in the panels on the left.  But the gas then collapses back into the halo 
sooner than in the \citet{jet13a} run, with fallback approaching the center by 
$\sim$ 25 Myr.  We note that the $\sim$ 50 km s$^{-1}$ gas velocities at radii 
greater than 500 pc are from cosmological infall and virialization.  They are 
steady throughout the run because accretion continues throughout the simulation.

\begin{figure}
\plotone{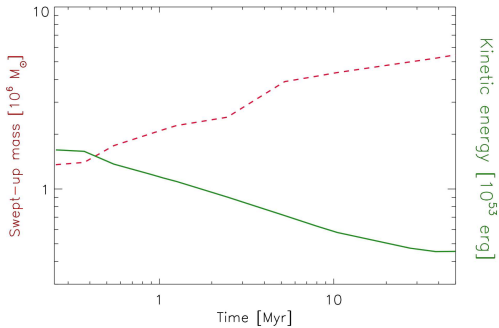} 
\caption{The mass swept up by the shock (\textit{red dashed line}) and the 
total kinetic energy of the flow (\textit{green solid line}) as a function of time.}
\vspace{0.1in}
\label{fig:KE}
\end{figure}

\begin{figure*}
\plotone{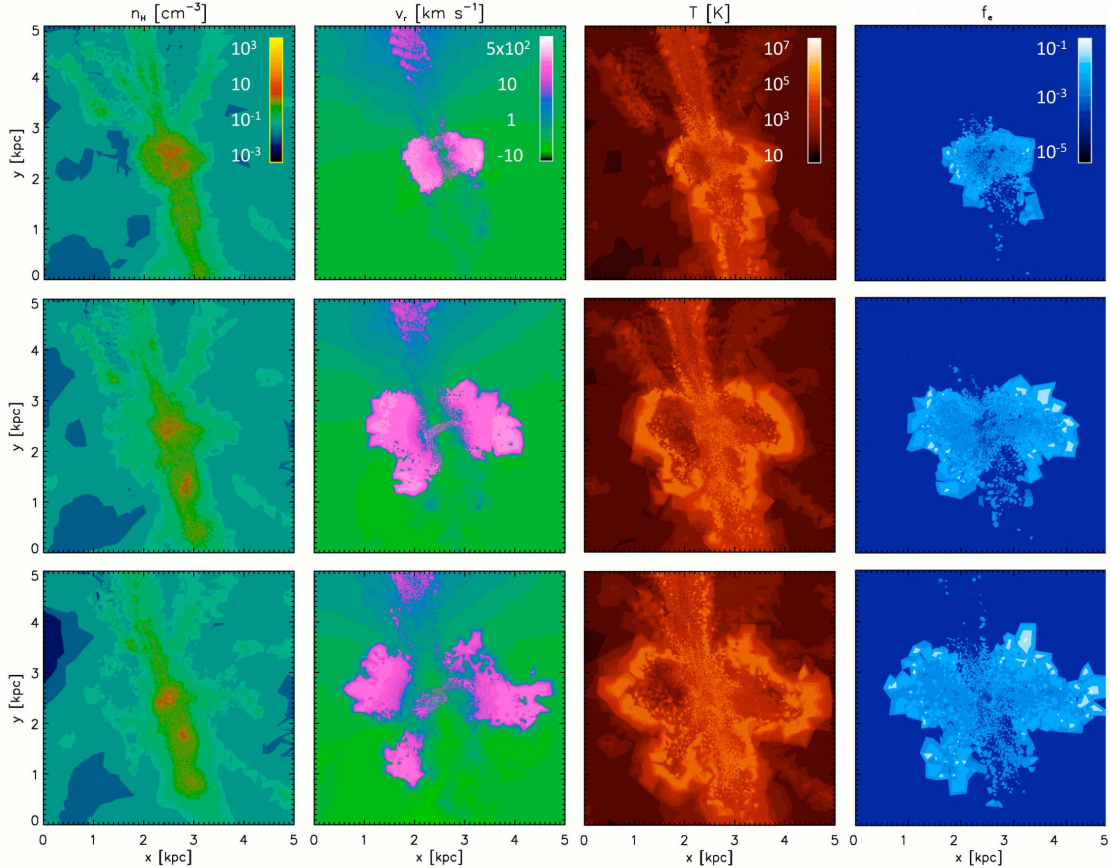} 
\caption{The SN at 10 Myr (\textit{top}), 25 Myr (\textit{center}) and 50 Myr 
(\textit{bottom}).  Left to right:  projections of H number density, radial gas 
velocity, temperature and free electron fraction.}
\vspace{0.1in}
\label{fig:proj}
\end{figure*}

As shown in Fig.~\ref{fig:KE}, the ejecta has lost 95\% of its original kinetic energy
by the time it is initialized in GADGET, but, as we show later, it still has enough 
linear momentum to drive gas in the halo out to its virial radius.  Line emission and 
expansion in the dark matter potential of the halo reduce the kinetic energy to 4.5 
$\times$ 10$^{52}$ erg by 25 Myr.  The ejecta displaces about the same amount 
of gas as in the \citet{jet13a} models ($\sim$ 10$^7$ \Ms) but to much smaller radii, 
2 kpc rather than 8 - 10 kpc.  We show the evolution of the protogalaxy in 
Fig.~\ref{fig:proj}.  Even with blowout into low-density voids, most of the gas is still 
confined to within $\sim$ 2 kpc of the center of the halo, unlike in \citet{jet13a} in 
which the ejecta reaches radii of 10 kpc, perhaps enriching nearby halos.  As seen 
in the velocity projections, gas preferentially propagates into the voids but is 
essentially halted along filaments by cosmological inflows. The SN expands to 
somewhat larger radii on average in this cosmological simulation than in the 
ZEUS-MP runs in \citet{wet13a} because of blowout into voids, which cannot be 
captured in 1D.

One physical process we do not capture in ZEUS-MP or GADGET is the dynamical
decoupling of ions from electrons in the shock that may occur 5 - 10 yr after the SN.  
If these two components do decouple the shock cools less efficiently because the
entropy generated by the shock, which is deposited in the ions, cannot be radiated
away efficiently by the electrons.  This phenomenon reduces cooling in galactic SN 
remnants in relatively diffuse ISM densities at intermediate times.  Electron-ion
decoupling does not affect the explosions of supermassive stars in diffuse regions
in line-cooled protogalaxies \citep{jet13a} because such explosions unbind baryons
from the halo anyway.  It could allow explosions in the dense regions studied here 
to grow to larger radii, but it is unclear for how long cooling would be curtailed.
Future improvements to RAGE, such as 3-temperature physics in which photons,
electrons and ions can be evolved with distinct temperatures, could address these 
issues.  The cooling found in our simulations should therefore be considered to be
upper limits for now.  

\subsection{Chemical Enrichment}

\begin{figure*}
\plotone{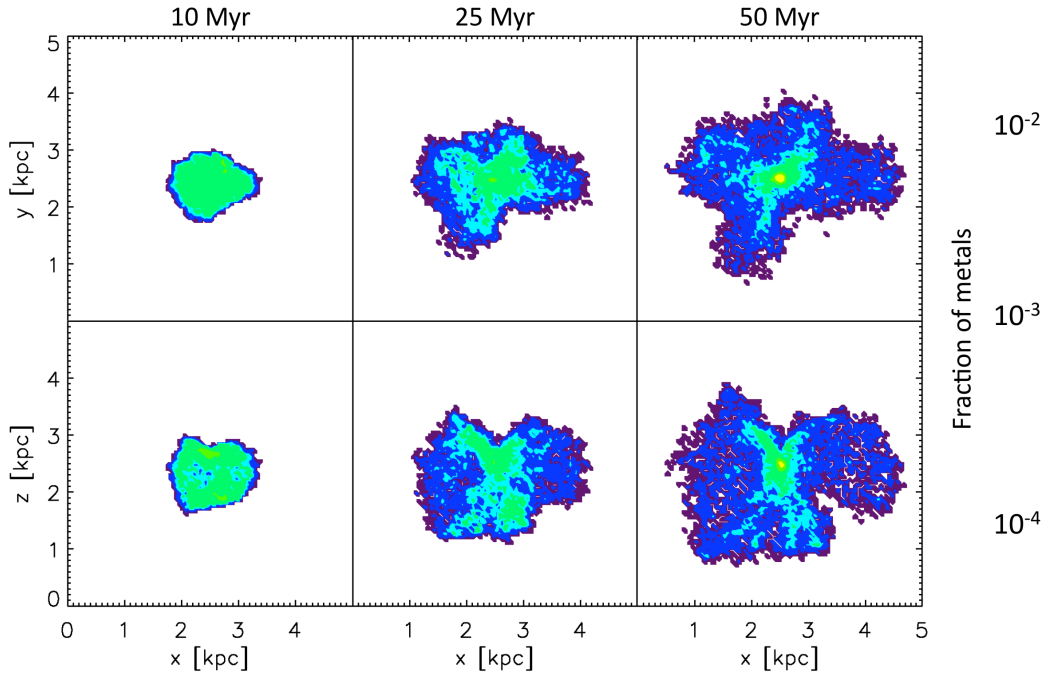} 
\caption{Projections of metallicity along the y- (\textit{top}) and the z-axes
(\textit{bottom}) at 10 Myr (\textit{left}), 25 Myr (\textit{center}) and 50 Myr 
(\textit{right}).}
\vspace{0.1in}
\label{fig:Z1}
\end{figure*}

As with Pop III PI SNe, about half of the mass of the supermassive SN is blown out 
into the galaxy in the form of heavy elements, $\sim$ 23,000 \Ms.  If the explosion 
occurs in a dense environment and drives turbulence, these metals can efficiently 
mix with gas in the halo, enhancing its cooling rates and altering the mass scales 
on which it fragments and forms new stars \citep[e.g.,][]{bcl01,ss06,schn06}, even 
in the presence of strong LW backgrounds \citep{osh08}.  We show projections of 
metallicity at 10, 25 and 50 Myr in Fig.~\ref{fig:Z1}.  The ejecta bubble is mostly 
uniform in metallicity when it reaches the virial radius but then some heavy elements 
continue out into the low-density voids while the rest begin to fall back into the halo.  
Later, as the collapse of ejecta back into the halo accelerates, inhomogeneities in 
metallicity become more pronounced.  As Fig.~\ref{fig:Z1} shows, metals propagate 
the least distance along the filaments (at whose intersection the halo originally 
formed).

\begin{figure}
\plotone{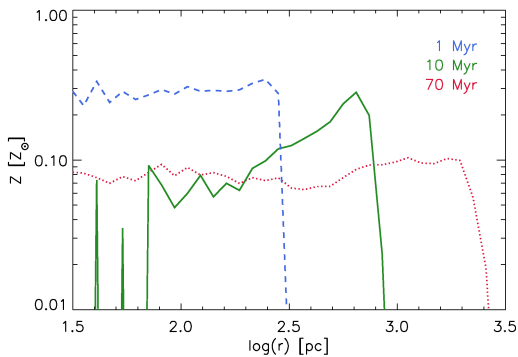}
\caption{Spherical average of the metallicity $Z$ of the gas (in units of solar metallicity 
\Zs) as a function of radius at 1 Myr (\textit{blue dashed line}), 10 Myr (\textit{green 
solid line}) and 70 Myr (\textit{red dotted line}).}
\vspace{0.1in}
\label{fig:Z2}
\end{figure}

We plot the evolution of the metallicity of the halo as a function of radius at 1, 10 
and 70 Myr in Fig.~\ref{fig:Z2}.  By 10 Myr the SN has essentially driven all the 
metals from the central 50 pc of the halo.  But, as shown in the bottom row of 
Fig.~\ref{fig:gas}, fallback has begun by $\sim$ 25 Myr, raising the metallicity $Z$ 
of the protogalaxy to 0.1 - 0.2 \Zs\ at $r \gtrsim$ 50 pc compared to 0.05 - 0.1 
\Zs\ at $r \gtrsim$ 100 pc in Fig.~5 of \citet{jet13a}.  It is clear that early fallback 
and the relative confinement of the ejecta in this run drives the interior of the 
protogalaxy to metallicities that are twice those of the explosions in \HII\ regions 
in \citet{jet13a}.  At 70 Myr the entire halo has been enriched to a metallicity $Z 
\sim$ 0.1 \Zs\ compared to $Z \sim$ 0.05 \Zs\ in \citet{jet13a}.  There were still a 
few metal particles less than 50 pc from the center of the halo at 10 Myr whose 
outward propagation was slowed by strong filamentary inflows along a few lines 
of sight.  These particles are visible as the two spikes in metallicity at radii of 
$\lesssim$ 50 pc in Fig.~\ref{fig:Z2}.

\begin{figure*}
\plotone{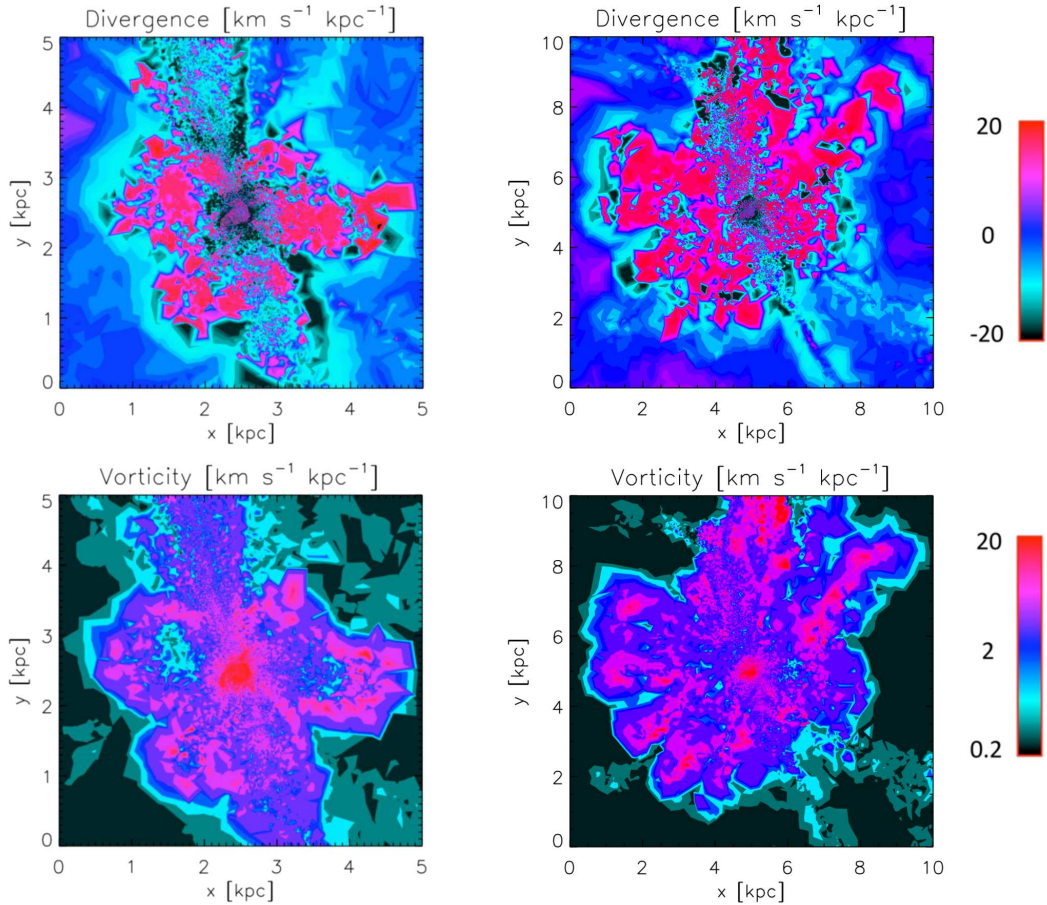} 
\caption{Divergence and vorticity in the protogalaxy due to supermassive explosions
in dense regions (\textit{left}) and diffuse \HII\ regions (\textit{right}).  The divergence 
($| \nabla \cdotp \vec{\bf{v}}|$, \textit{top row}) marks regions of rapidly expanding 
flow and vorticity ($|\nabla \times \vec{\bf{v}}|$, \textit{bottom row}) traces shear flows 
that probably drive turbulence.}
\vspace{0.1in}
\label{fig:curl}
\end{figure*}

\citet{wet13a} speculated that massive fallback into the halo would not only distribute
metals throughout its interior:  it, together with cosmological inflows, would also drive 
turbulence that would efficiently mix the metals with baryons in the protogalaxy and 
radically alter their cooling properties.  However, their 1D models could not simulate 
these processes.  We show the vorticity $|\nabla \times \vec{\bf{v}}|$ due to expansion 
and fallback in a slice through the center of the halo at 50 Myr for this run and \citet{
jet13a} in Fig.~\ref{fig:curl}.  Vorticity is a measure of the shear flows that drive 
Kelvin-Helmholtz instabilities that in turn energize turbulent cascades in the gas.  
Wherever it is strong, turbulence and efficient mixing are likely to follow.  As shown in 
the bottom row of Fig.~\ref{fig:curl}, both explosions exhibit similar degrees of vorticity, 
although it is greater at the center of the protogalaxy in the present run where early 
fallback and accretion collide and churn the gas to a greater degree than in \citet{
jet13a}.  In both cases it is anti-correlated with blowout into low-density voids, which is 
characterized by rapid expansion without much shear.  These flows are marked by the 
regions of large divergence $| \nabla \cdotp \vec{\bf{v}}|$ shown in the top row of 
Fig.~\ref{fig:curl}.  

The large vorticities in both explosions suggests that turbulence will enrich most of 
the baryons in the protogalaxy with metals (and perhaps dust).  Star formation and 
a prompt rollover from a Pop III to a Pop II initial mass function (IMF) may follow 
throughout the interior of the halo. Although such explosions may be rare events in 
the early universe, their host galaxies would be easily distinguished from their less 
rapidly evolving neighbors by their large star formation rates and by their distinct 
spectral energy distributions (SEDs), which would be almost entirely due to Pop II 
stars.  

\subsection{Fallback and SMBH Seed Growth}

\begin{figure}
\plotone{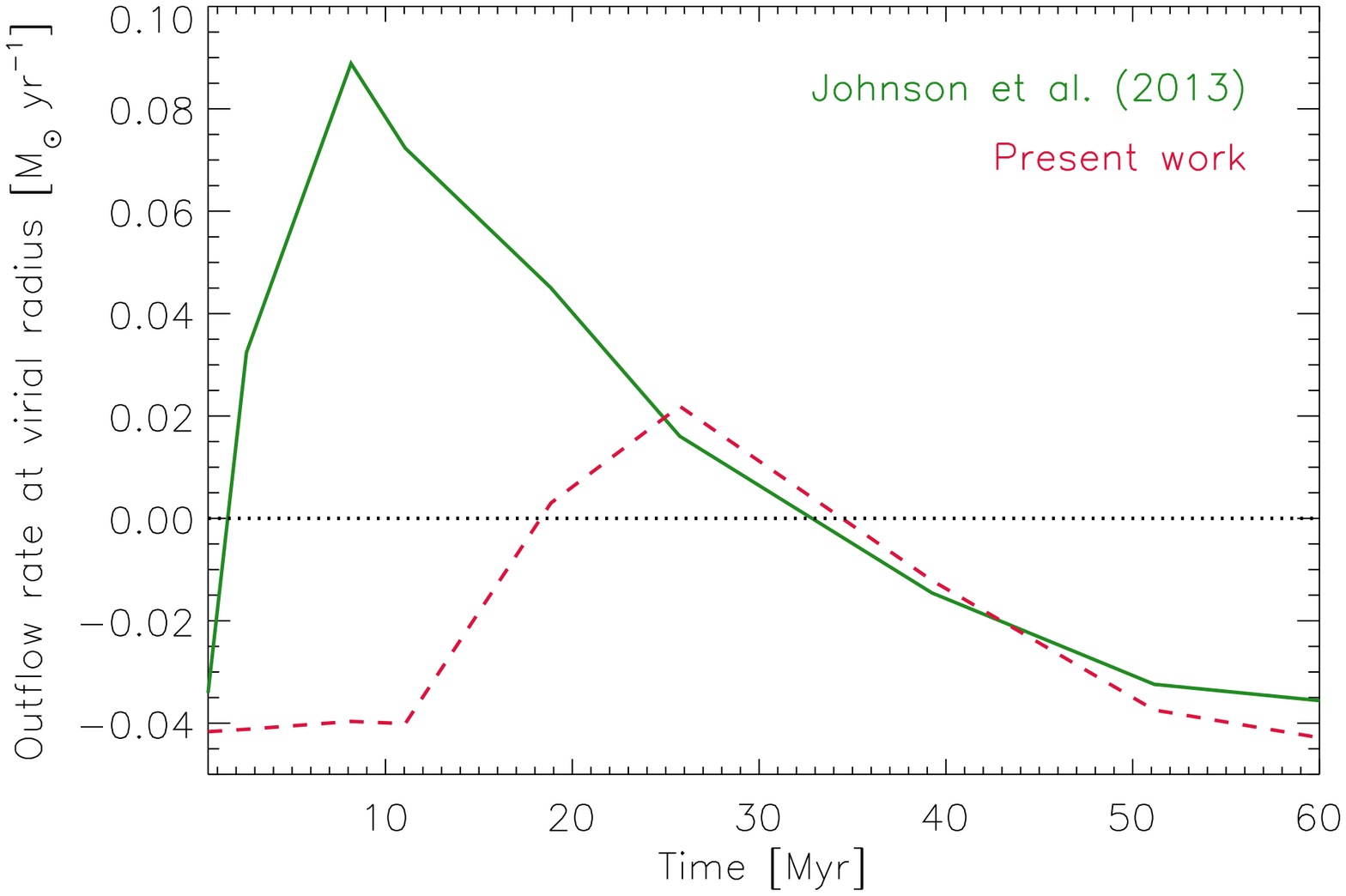}
\caption{Outflow and fallback rates in the protogalaxy as a function of time for 
explosions in diffuse \HII\ regions (\textit{green line}) and dense regions (\textit{red 
dashed line}).}
\vspace{0.1in}
\label{fig:fb}
\end{figure}

Outflow and fallback through a spherical boundary at 1 kpc, the virial radius of the
halo, are shown for this run and for \citet{jet13a} in Fig.~\ref{fig:fb}. Differences in the 
overall dynamics between SNe in dense accretion envelopes and diffuse \HII\ regions 
are abundantly clear. First, blowout happens sooner in low density environments than 
in heavy infall, in which the shock does not reach the virial radius and reverse 
accretion onto the halo until $\sim$ 10 Myr. Peak outflow rates are also much higher in 
\citet{jet13a} due to lower energy losses and breakout into low density voids.  Fallback 
is heavier in our run at late times because little of the ejecta is unbound from the halo, 
unlike \citet{jet13a} in which metals blown out into voids only return after a fraction of a 
Hubble time.  It might be thought from the relative positions of the two peaks in outflow 
that fallback begins earlier in \citet{jet13a} than in this run.  In reality, as noted above, 
fallback begins earlier in explosions in dense regions. In 
these cases much of the ejecta never reaches 1 kpc and does not appear in the plot.  
These metals fall back to the center of the halo well before ejecta in explosions in low
densities returns to the halo \citep[compare Fig.~\ref{fig:gas} in this study to Fig.~2 
in][]{jet13a}. 
 
Steady cosmological inflow from filaments is evident in the present run until $\sim$ 10 
Myr, after which the SN reverses them.  Comparing this plot to Figs.~6 and 7b in \citet{
wet13a}, it is clear that cosmological flows somewhat dampen the massive fallback of 
1D models, averaging them out over time. This happens in part because metals along 
some lines of sight fall back into the halo sooner as they encounter inflow along dense 
filaments.  As shown in the left panels of Fig.~\ref{fig:curl}, accretion and fallback also 
drive turbulent motions in the halo that partially support the gas against collapse.  
However, fallback eventually drives central accretion rates to values greater than those 
originally due to filaments alone, as shown in the slope of the red plot after 55 Myr.  

Fallback in both explosions eventually results in flows capable of rapidly growing SMBH 
seeds at the center of the protogalaxy.  How x-rays from nascent black holes regulate 
these infall rates ($\gtrsim$ 0.04 \Ms\ yr$^{-1}$) remains to be determined.  
Nevertheless, it is probable that baryon collapse rates of this magnitude could drive 
episodes of super-Eddington accretion by central BHs.  We note that fragmentation of 
the atomically-cooled disk at the center of the halo almost certainly leads to multiple 
supermassive clumps \citep[see Fig.~1 of][]{wet12d}.  Supermassive stars and SMBH 
seeds therefore probably coexist at the center of the protogalaxy, and massive BHs 
would remain after the explosion of one of the stars.
 
\section{Discussion and Conclusion}

In \citet{jet13a} and this paper, we have considered the two extremes for 
supermassive Pop III SNe in line-cooled protogalaxies:  explosions in diffuse 
regions, like an \HII\ region of the star, and explosions in dense envelopes, 
like those that gave birth to the star.  We find that SNe in dense protogalactic 
cores do not grow to radii much larger than that of the halo, unlike explosions 
in low-density regions whose metals can envelop and enrich nearby galaxies.  
In reality, explosions of supermassive Pop III stars occur in regions that fall 
somewhere in between these two extremes, depending on the rates and 
geometry of accretion and whether or not ionizing UV radiation from the 
progenitor breaks out of its accretion envelope \citep{jlj12a}.  These 
spectacular events are likely accompanied by intense starbursts and rapid 
growth of SMBH seeds driven by massive fallback and prompt chemical 
enrichment of baryons in the halo.

Supermassive Pop III SNe can be detected on multiple, disparate timescales
after they occur, beginning with the prompt NIR emission from the initial blast.
\citet{wet12d} find that explosions in dense envelopes are much brighter in 
the NIR than SNe in diffuse regions and can be detected at $z \sim$ 15 - 20
by WFIRST, WISH and \textit{JWST} and at $z \sim 10 - 15$ by \textit{Euclid}.
This point is important because all-sky surveys by missions such as WFIRST, 
WISH and \textit{Euclid} could discover these explosions in spite of their small
numbers.  Later, on timescales of $\sim$ 10$^3$ yr, the SN remnant becomes 
a strong synchrotron source that is visible in the radio to eVLA, eMERLIN, 
ASKAP and the Square Kilometer Array (SKA) at $z \gtrsim$ 15 \citep{met12a,
wet13a}.  In contrast to the NIR, this emission is much brighter for explosions 
in \HII\ regions because of the higher shock temperatures and it has a profile 
that easily distinguishes it from less energetic Pop III SNe.

Later, on times of 10 - 20 Myr, fallback may activate emission by SMBH seeds 
in the halo that could be detected in the NIR by successors to \textit{Swift} such 
as the Joint Astrophysics Nascent Universe Satellite \citep[JANUS;][]{Roming08,
Burrows10}, Lobster, or EXIST.  At about this time chemical enrichment and 
cooling may also trigger starbursts that could be detected in the NIR by 
\textit{JWST}, distinguishing these halos from their less luminous neighbors.  
Testing the prospects for rapid star formation and SMBH seed growth in line
cooled halos after such explosions requires numerical simulations with metal 
and dust mixing and cooling together with x-ray feedback by the central BH that 
are now under development.  We note that these explosions in principle could 
also be found in the CMB via the Sunyayev-Zel'dovich effect, but the likelier 
candidates for such detections are SNe in diffuse \HII\ regions that enclose and 
upscatter greater numbers of CMB photons \citep[e.g.,][]{oh03,wet08a}.

The debris of these ancient explosions could also be found in the 
atmospheres of dim metal-poor stars in the Galactic halo today \citep[e.g.,][]{
bc05,fet05,Cayrel2004,Lai2008,fb12}.  Gas enriched to the metallicities in our
simulation is expected to fragment into stars $\lesssim$ 0.8 \Ms\ that may still 
exist, particularly if they formed in dust \citep{schn06}.  The chemical 
signatures of supermassive SNe can be distinguished from those of Pop III PI 
SNe because they create very little \Ni, unlike PI SNe that produce iron-group 
elements \citep{hw02}.  However, like PI SNe supermassive SNe make few 
r-process and s-process elements, so their chemical fingerprint can also be 
differentiated from those of 15 - 40 \Ms\ Pop III core-collapse SNe.  These 
abundance patterns may not have been detected yet because the stars that 
bear them were enriched to metallicities above those targeted by surveys of 
metal-poor stars to date \citep{karl08} \citep[see also][for tentative evidence 
of Pop III PI SNe in the fossil abundance record]{cooke11,ren12}.

Besides marking the sites of formation of SMBHs on the sky, supermassive 
Pop III SNe would also reveal the location of their nurseries:  line-cooled halos
and the nearby protogalaxies that saturate them with strong LW fluxes 
\citep[e.g.,][]{dijkstra08,agarw12}.  These rapidly forming young galaxies 
could then be selected for spectroscopic followup by \textit{JWST}.  
Detections of the most energetic thermonuclear explosions in the universe
at $z \gtrsim$ 15 by the next generation of telescopes may finally reveal the
origins of SMBHs.

\acknowledgments

We thank the anonymous referee, whose comments improved the quality of this 
paper.  JLJ and JS were supported by LANL LDRD Director's Fellowships.  DJW 
acknowledges support from the Baden-W\"{u}rttemberg-Stiftung by contract 
research via the programme Internationale Spitzenforschung II 
(grant P- LS-SPII/18).  AH was supported by the US DOE Program for Scientific 
Discovery through Advanced Computing (SciDAC; DE-FC02-09ER41618), by the 
US Department of Energy under grant DE-FG02-87ER40328, by the Joint Institute 
for Nuclear Astrophysics (JINA; NSF grant PHY08-22648 and PHY110-2511).  AH 
also acknowledges support by an ARC Future Fellowship (FT120100363) and a 
Monash University Larkins Fellowship. Work at LANL was done under the auspices 
of the National Nuclear Security Administration of the U.S. Department of Energy at 
Los Alamos National Laboratory under Contract DE-AC52-06NA25396. All GADGET 
and ZEUS-MP simulations were performed on Institutional Computing platforms at 
LANL (Mustang and Pinto).

\bibliographystyle{apj}
\bibliography{refs}

\end{document}